\documentclass{ws-ijmpa}

\usepackage{graphicx}
\usepackage{bm}
\usepackage{psfig}  
\usepackage{epsfig}

\begin{document}

\markboth{Ivan Vitev}
{Probing the phases of QCD in ultra-relativistic nuclear collisions}

%
\catchline{}{}{}{}{}
%

\title{ PROBING THE PHASES OF QCD \\ IN ULTRA-RELATIVISTIC NUCLEAR 
COLLISIONS
}

\author{\footnotesize IVAN VITEV }

\address{Los Alamos National Laboratory, Theory Division and 
Physics Division \\
Mail Stop H846, Los Alamos, NM 87545, USA
}

\maketitle

\pub{Received 25 November 2004}{}

\begin{abstract}

The status of RHIC theory and phenomenology is reviewed with an 
emphasis on the indications for the creation of a new deconfined 
state of matter. The critical role of high energy nuclear physics 
in the development of theoretical tools that address various 
aspects of the QCD many body dynamics is highlighted. The perspectives 
for studying nuclear matter under even more extreme conditions 
at the LHC and the overlap with high energy physics 
is discussed.

\end{abstract}

\section{Introduction}

Quantum Chromodynamics and asymptotic freedom\cite{Gross:1974cs} 
predict the existence of a new state of matter\cite{Collins:1974ky}, 
the quark-gluon plasma (QGP), at exceedingly high temperatures and 
energy densities similar to the ones that characterized the first 
few microseconds after the Big Bang. The quest for such a deconfined 
state of QCD has led to the highly successful program at the Relativistic 
Heavy Ion Collider (RHIC) and will be an integral part of the program 
at the Large Hadron Collider (LHC).  Recent theoretical developments 
have improved our understanding of the complex many-body dynamics in 
high energy nuclear collisions. Comparison between data and theory 
is suggestive of the creation of a deconfined state of QCD with energy 
density on the order of 100 times normal nuclear matter density. 
For a review of the RHIC experimental results see\cite{Nagle}.

\section{Soft and Intermediate $p_T$ Hadrons}

An economical description of the bulk particle production in $A+A$ reactions
can be achieved in the framework of the thermal model.
All measured  particle ratios at  midrapidity from the  
$\sqrt{s_{NN}}=130, \, 200$~GeV $Au+Au$ runs 
at RHIC are well reproduced  with 
$T_f \approx T_c \simeq 175$~MeV and $\mu_B = 40,\, 30$~MeV, 
respectively\cite{Braun-Munzinger:2001ip}. Such approach, while 
instructive,  does not carry information about the dynamical evolution 
prior to freeze-out. Constraining the initial conditions, relevant 
at the early stages of relativistic heavy ion collisions, requires 
the development and application of microscopic models.

{\it Relativistic hydrodynamics.}
Differential particle distributions  at low $p_T$ can be calculated
in the framework of relativistic hydrodynamics\cite{Kolb:2003dz}, 
which assumes local thermal equilibrium and solves the energy-momentum 
and current conservation
\begin{equation}
\partial_\mu T^{\mu\nu}(x) = 0\, , \qquad  \partial_\mu j_i^{\mu}(x) = 0 \;. 
\label{hydro}
\end{equation}
Additional constraints, necessary to determine the 
system Eq.~(\ref{hydro}), include, for example, ideal hydrodynamics 
and equation of state $p=p(\epsilon)$ or a boost invariant Bjorken model.  
Good description of the  $p_T \leq 2 - 3$~GeV 
hadron spectra for $\pi^\pm,\,  K^\pm,\,  p (\bar{p}), \, \Lambda
(\bar{\Lambda}) ,  \, \Omega \, , \cdots$ can be obtained. However, the 
elliptic flow $v_2$ of massive hadrons  was shown to be much 
more sensitive to the choice of equation of state 
(EOS) with first order phase transition (preferred) versus hadronic 
EOS\cite{Huovinen:2001wn}. 
Traces of early partonic thermalization, as suggested by hydrodynamics, 
can also be found in the in the mean $p_T$  fluctuations 
versus centrality\cite{Gavin:2003cb}.

Recent hydrodynamic simulations have employed initial conditions 
motivated by gluon saturation phenomenology\cite{Hirano:2004rs}.  
In  $A+A$ reaction this approach alleviates the problem with the 
transverse energy and the parton number in such models. However, 
it leaves an open question  for $p+A$ reactions where the hydrodynamic 
description is not  applicable.

{\it Parton Coalescence.}
Another microscopic approach to low and moderate $p_T$ hadroproduction 
is the covariant Boltzmann transport\cite{Molnar:2000jh}. 
The theory was shown to be applicable  not only in $A+A$  but also 
in $p+A$ reactions\cite{Lin:2003ah} 
and has provided a good description of the rapidity density of hadrons 
$dN^{ch}/dy$ and their transverse momentum distributions. 
The output of a partonic transport can be incorporated in 
the coalescence/recombination models that convolute the quark Wigner 
functions with the meson and baryon wavefunctions\cite{Fries:2004ej}. 
These predict scaling properties of the elliptic flow 
$v_2$\cite{Molnar:2003ff}
\begin{equation}  
v_2^M(p_T) = 2 v_2^p(p_T/2)\, , \qquad  v_2^B(p_T) = 3 v_2^p(p_T/3)\;  
\end{equation}  
at intermediate transverse momenta and enhanced baryon to meson 
ratios such as $p(\bar{p})/\pi^\pm$ and $\Lambda(\bar{\Lambda})/K^\mp$.
The energy loss of quarks and gluons plays an important role in the
generation of the elliptic flow at the partonic level\cite{Gyulassy:2000gk}.
It also suppresses the perturbative high $p_T$ hadron production 
and thus extends the region of possible non-perturbative contribution for 
baryons\cite{Vitev:2001zn}. 
Coalescence models have been used to calculate the  expected charm meson 
transverse momentum distributions and the charm meson and baryon 
elliptic flow\cite{Greco:2003vf}.

\section{The Perturbative QCD Factorization Approach}

In the nuclear environment,  modifications to the 
QCD factorization approach\cite{Collins:gx} arise from the 
elastic, inelastic and coherent multiple scattering. These can 
be systematically incorporated in to the perturbative 
formalism\cite{Vitev:2004kd}. For $\ell+A$, $p+A$ and
$A+A$ reactions their effect on  experimental observables is
summarized in Table~1. The relative importance of these distinct 
contributions in the different kinematic and density regimes is 
yet to be determined. 
\begin{table}[b!]
\label{table1}
\tbl{Effect of elastic, inelastic and coherent multiple scattering on 
the transverse momentum distribution of hadrons in the 
perturbative regime.}
{\begin{tabular}{@{}ll@{}} 
\toprule
{\bf Type of  scattering}  \ \ \ \  &
{\bf Transverse momentum dependence of the nuclear effect} \\[2ex]
\toprule
{\it Elastic}  &  Suppression at low $p_T$, enhancement at moderate $p_T$, 
disappears at high $p_T$ \\
{\it Inelastic} & Suppression at all $p_T$, weak $p_T$ dependence, persists at 
high $p_T$   \\
{\it Coherent}  & Suppression at low $p_T$, disappears at high $p_T$ \\
\botrule
\end{tabular}}
\end{table}
Elastic scattering leads to a Cronin enhancement and a small broadening 
of the away side correlation function $C(\Delta \phi) = 
(1/N_{trig})dN^{h_1h_2}/d\Delta \phi$ . These have 
been discussed in\cite{Vitev:2003xu}.


\begin{figure}
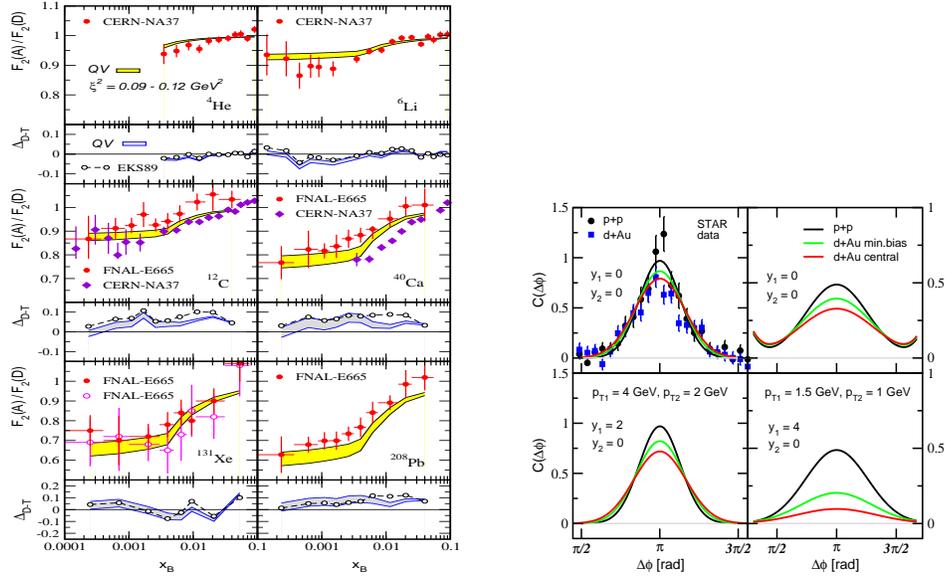

\vspace*{.4cm}
\centerline{\psfig{file=fig2.eps,width=6.cm,height=7.6cm}
\hspace*{.7cm} \psfig{file=Fig3.eps,width=5.5cm}}
\vspace*{8pt}
\caption{Left panel from\protect\cite{Qiu:2003vd}:
resummed QCD power corrections to nuclear shadowing in $\ell +A$, manifest 
in $F_2(A)/F_2(D)$. Right panel from\protect\cite{Qiu:2004da}:
modification of the away side correlation function 
$ C(\Delta \phi)$
versus rapidity, centrality  and {$p_T$} from  higher twist.} 
\end{figure}

{\it Coherent power corrections.}
A class of corrections that can be naturally incorporated
in  the perturbative QCD factorization approach is associated 
with the power suppressed $\sim 1/Q^n,\; n \geq 1$ contributions.
The higher twist terms are typically  neglected in reactions 
with ``elementary'' nucleons for $Q^2 \geq 1$~GeV$^2$. 
However, in the presence of nuclear  matter such corrections 
can be enhanced by the large nuclear size $\sim A^{1/3}$.

The effective longitudinal interaction length probed by the virtual meson
of momentum $q^\mu$ is characterized by $1/xP$. If the momentum fraction 
of an active initial-state parton $x \ll x_c=1/2m_N r_0\sim 0.1$  
with nucleon mass $m_N$ and radius $r_0$, it could cover several  
Lorentz contracted nucleons of longitudinal size $\sim 2r_0 (m_N/P)$
in a large nucleus.
Each of the soft interactions is characterized by a 
scale of power correction per nucleon
$\xi^2\approx \frac{3\pi\alpha_s(Q)}{8r_0^2}
\langle p|\hat{F}^2|p\rangle$
with the matrix element  $\langle p|\hat{F}^2|p\rangle = 
\frac{1}{2} \lim_{x\rightarrow 0}xG(x,Q^2)$\cite{Qiu:2003vd}. 
The multiple final state scattering of the struck quark 
generates a dynamical parton mass $m_{\rm dyn}^2 =\xi^2 (A^{1/3} -1) $  
and a consequent rescaling in the value of Bjorken-$x$\cite{Qiu:2003vd}:
\begin{equation}
\hspace*{-1cm} x_B \rightarrow x_B + x_B \frac{M_q^2}{Q^2} +  
x_B \frac{\xi^2 (A^{1/3} -1)}{Q^2} = 
x_B \left( 1 + \frac{M^2_q + m_{\rm dyn}^2 }{Q^2}    \right)  \;.  
\label{shift} 
\end{equation}     
In Eq. (\ref{shift}) $M_q$  is  the physical mass of the quark
in the final state.

In Ref.~\cite{deFlorian:2003qf} next-to-leading order global analysis
of the nuclear parton distributions (nPDFs) was shown to strongly 
disfavor more than $5\%$ gluon shadowing. This  result also
suggests that the observed attenuation in the structure functions 
is a result of multiple final state 
scattering since any dynamical mechanism will predict an effect 
twice as large for gluons when compared to quarks, 
$ \xi^2 \rightarrow C_A/C_F \xi^2$.

Application to nuclear shadowing in the DIS structure 
function $F_2(A)$ is shown in  the  left panel of Fig.~1. 
The coherent nuclear enhanced power  corrections in single and 
double inclusive  hadron production  in $p+A$ reactions have 
been evaluated in\cite{Qiu:2004da}. The apparent suppression 
of the  away side correlation function 
$ C(\Delta \phi) = (1/N_{trig})dN^{h_1h_2}/d\Delta \phi$ is shown 
in the right hand side of 
Figure~1. As seen from Eq. (\ref{shift}), these modifications disappear 
at high $p_T$. It will be instructive to investigate  the open charm 
production\cite{ming} including resummed power corrections. 
Additional discussion of nuclear shadowing and 
antishadowing in the light front approach can be 
found in\cite{Brodsky:2004er}.

{\it Non-Abelian energy loss.}
The most efficient mechanism that modifies the large 
transverse momentum hadroproduction
in the presence of a hot and dense quark-gluon plasma is the medium
induced gluon bremsstrahlung\cite{Gyulassy:2003mc}. It leads to strong 
suppression of high-$p_T$ particles, known as jet quenching.
Non-Abelian energy loss was also shown to be an 
effective attenuation mechanism  in cold nuclear 
matter\cite{Wang:2002ri}.

The qualitative behavior of the energy loss  as a function 
of the density and the size of the system can be calculated 
using the GLV approach\cite{Gyulassy:2000er}. 
To first order in opacity  for static and 1+1D Bjorken expanding 
plasmas
\begin{eqnarray}
\langle \Delta E \rangle &\approx&  
 \left\{ \begin{array}{ll}  
\frac{9 C_R \pi \alpha_s^3}{8} \, \rho^g \langle L \rangle^2 
\, \ln \frac{2E}{\mu^2 \langle L \rangle } 
 \, ,  &  \qquad {\rm static} \\[1ex]
 \frac{9  C_R \pi \alpha_s^3}{4} 
\frac{1}{A_\perp}  \frac{dN^{g}}{dy} \langle L  \rangle 
\,   \ln \frac{2E}{\mu^2 \langle L \rangle}  
 \, , & \qquad 1+1D  
\end{array}  \right.   \;.  
\label{deltae}
\end{eqnarray}
Application\cite{Vitev:2004gn} of the QCD theory of energy 
loss to the single inclusive pion suppression and the attenuation 
of the away side dijet correlations is shown in Fig.~2. These persist,
as shown by the calculation, to high $p_T$. For similar results 
see\cite{Wang:2004tt}.

Better experimental techniques may reveal the redistribution of 
the lost jet energy\cite{Pal:2003zf} in low frequency modes 
$\geq \omega_{pl}$. The quenching of charm and beauty 
mesons\cite{Djordjevic:2003zk} will provide a physical scale, 
$m_c$ and $m_b$, relative to which the parameters of the medium 
can be reliably extracted.

\begin{figure}
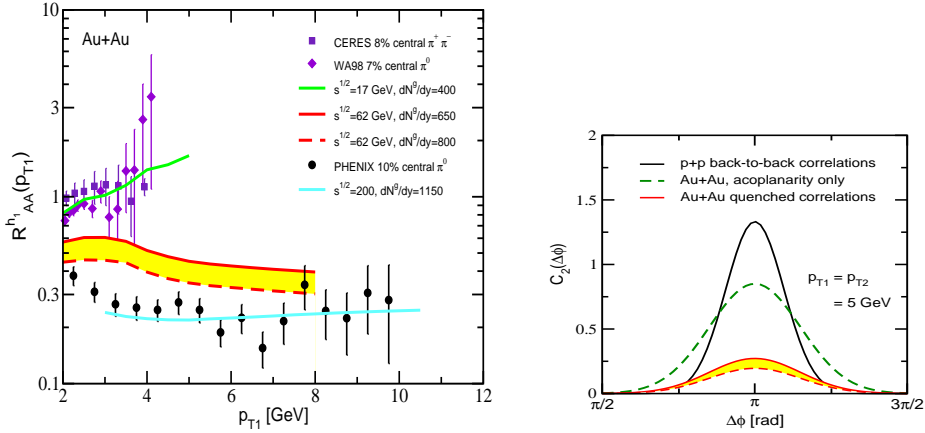

\centerline{\psfig{file=Fig2a.proc.eps,width=6.5cm,height=5.7cm}
\hspace*{.5cm} \psfig{file=Fig2b.proc.eps,width=5cm,height=4.cm }}
\caption{Left panel from\protect\cite{Vitev:2004gn}:
nuclear modification factor $R_{AA}(p_T)$ at $\sqrt{s_{NN}} = 
17, 62$ and $200$~GeV.  
Right panel from\protect\cite{Vitev:2004gn}:
modification of the away side correlation function 
$ C(\Delta \phi)$ from energy loss  at high $p_T$.} 
\vspace*{8pt}
\end{figure}

\section{Conclusions}

The complex dynamics of ultrarelativistic heavy ion reactions 
invites a variety of theoretical approaches, applicable in  
the different kinematic and density regimes, in the description
of experimental observables. Consistent application of the models 
is required to better understand their strengths and limitation 
and to  disentangle the distinct nuclear effects.   
At present, combined results from relativistic hydrodynamics 
and jet tomography  suggest the creation of QGP at RHIC with 
initial energy density  $\epsilon \simeq 15 - 20$~GeV/fm$^3$. 
The interactions in the plasma, however,  remain strong and are 
consistent with the lattice expectation of $\alpha_s 
\simeq 0.5$\cite{alphas}.

The upcoming RHIC runs and the LHC program will ensure the dominance 
of high $p_T$ and jet physics\cite{Collins:1981uk} and will provide 
precision $A+A$ data at $\sqrt{s_{NN}} = 200\, - \, 5500$~GeV\cite{jets}. 
This opens new possibilities for studying photon and dilepton
tagged jets, heavy flavor and multiparticle correlations.  
For cold nuclear matter, a low energy $p(d)+A$ run at RHIC 
will be a critical step in clarifying the relative importance 
of the nuclear effects summarized  in Table~1.

{\it Acknowledgments.}
I thank the organizers for the invitation to give this talk.
This work is supported by the J.R.~Oppenheimer Fellowship 
of the Los Alamos National Laboratory and by the US Department 
of Energy.


\begin{thebibliography}{0}

\bibitem{Gross:1974cs}
D.~J.~Gross and F.~Wilczek,
Phys.\ Rev.\ D {\bf 9}, 980 (1974);
H.~D.~Politzer,
Phys.\ Rev.\ Lett.\  {\bf 30} (1973) 1346.



\bibitem{Collins:1974ky}
J.~C.~Collins and M.~J.~Perry,
Phys.\ Rev.\ Lett.\  {\bf 34}, 1353 (1975).


\bibitem{Nagle} 
J.~L. Nagle, nucl-ex/0411027, these proceedings.


\bibitem{Braun-Munzinger:2001ip}
P.~Braun-Munzinger {\it et al.}, 
Phys.\ Lett.\ B {\bf 518}, 41 (2001); 
D.~Magestro, $\sqrt{s_{NN}}=200$~GeV update thereafter.


\bibitem{Kolb:2003dz}
P.~F.~Kolb and U.~Heinz,
nucl-th/0305084;
P.~Huovinen,
nucl-th/0305064; 
references therein.

\bibitem{Huovinen:2001wn}
P.~Huovinen, P.~F.~Kolb and U.~W.~Heinz,
Nucl.\ Phys.\ A {\bf 698}, 475 (2002).


\bibitem{Gavin:2003cb}
S.~Gavin,
Phys.\ Rev.\ Lett.\  {\bf 92}, 162301 (2004);
M.~A.~Aziz and S.~Gavin,
Phys.\ Rev.\ C {\bf 70}, 034905 (2004).




\bibitem{Hirano:2004rs}
T.~Hirano and Y.~Nara,
Nucl.\ Phys.\ A {\bf 743}, 305 (2004).


%
\bibitem{Molnar:2000jh}
D.~Molnar and M.~Gyulassy,
Phys.\ Rev.\ C {\bf 62}, 054907 (2000);
B.~Zhang,
Comput.\ Phys.\ Commun.\  {\bf 109}, 193 (1998).



\bibitem{Lin:2003ah}
Z.~W.~Lin and C.~M.~Ko,
Phys.\ Rev.\ C {\bf 68}, 054904 (2003).




\bibitem{Fries:2004ej}
R.~J.~Fries,
J.\ Phys.\ G {\bf 30}, S853 (2004), refererences therein;



\bibitem{Molnar:2003ff}
D.~Molnar and S.~A.~Voloshin,
Phys.\ Rev.\ Lett.\  {\bf 91}, 092301 (2003).


\bibitem{Gyulassy:2000gk}
M.~Gyulassy, I.~Vitev and X.~N.~Wang,
Phys.\ Rev.\ Lett.\  {\bf 86}, 2537 (2001);
M.~Gyulassy {\em et al.}, 
Phys.\ Lett.\ B {\bf 526}, 301 (2002).


\bibitem{Vitev:2001zn}
I.~Vitev and M.~Gyulassy,
Phys.\ Rev.\ C {\bf 65}, 041902 (2002).



\bibitem{Greco:2003vf}
V.~Greco, C.~M.~Ko and R.~Rapp,
Phys.\ Lett.\ B {\bf 595}, 202 (2004);
Z.~W.~Lin and D.~Molnar,
Phys.\ Rev.\ C {\bf 68}, 044901 (2003).


\bibitem{Collins:gx}
J.~C.~Collins, D.~E.~Soper and G.~Sterman,
Adv.\ Ser.\ Direct.\ High Energy Phys.\  {\bf 5} (1988) 1;


\bibitem{Vitev:2004kd}
I.~Vitev, to appear in J.\ Phys.\ G,
hep-ph/0409297;  
hep-ph/0410045;
J.~W.~Qiu and I.~Vitev,
hep-ph/0410218.



\bibitem{Vitev:2003xu}
I.~Vitev,
Phys.\ Lett.\ B {\bf 562}, 36 (2003);
J.~W.~Qiu and I.~Vitev,
Phys.\ Lett.\ B {\bf 570}, 161 (2003);
M.~Gyulassy, P.~Levai and I.~Vitev,
Phys.\ Rev.\ D {\bf 66}, 014005 (2002).



\bibitem{Qiu:2003vd}
J.~W.~Qiu and I.~Vitev, Phys.\ Rev.\ Lett.\ in press,
hep-ph/0309094;
Phys.\ Lett.\ B {\bf 587}, 52 (2004).



\bibitem{deFlorian:2003qf}
D.~de Florian and R.~Sassot,
Phys.\ Rev.\ D {\bf 69}, 074028 (2004).




\bibitem{Qiu:2004da}
J.~W.~Qiu and I.~Vitev,
hep-ph/0405068.



\bibitem{ming}
M.~X.~Liu,
nucl-ex/0405034;
these proceedings.



\bibitem{Brodsky:2004er}
S.~J.~Brodsky,
hep-ph/0411056;
S.~J.~Brodsky and H.~J.~Lu,
Phys.\ Rev.\ Lett.\  {\bf 64}, 1342 (1990).



\bibitem{Gyulassy:2003mc}
M.~Gyulassy {\em et al.}, 
nucl-th/0302077;
R.~Baier, D.~Schiff and B.~G.~Zakharov,
Ann.\ Rev.\ Nucl.\ Part.\ Sci.\  {\bf 50}, 37 (2000).



\bibitem{Wang:2002ri}
E.~Wang and X.~N.~Wang,
Phys.\ Rev.\ Lett.\  {\bf 89}, 162301 (2002).




\bibitem{Vitev:2004gn}
I.~Vitev,
nucl-th/0404052;
I.~Vitev and M.~Gyulassy,
Phys.\ Rev.\ Lett.\  {\bf 89}, 252301 (2002);



\bibitem{Gyulassy:2000er}
M.~Gyulassy, P.~Levai and I.~Vitev,
Nucl.\ Phys.\ B {\bf 594}, 371 (2001);
Phys.\ Rev.\ Lett.\  {\bf 85}, 5535 (2000);
Nucl.\ Phys.\ B {\bf 571}, 197 (2000).


\bibitem{Wang:2004tt}
Q.~Wang and X.~N.~Wang,
nucl-th/0410049; 
A.~Adil and M.~Gyulassy,
Phys.\ Lett.\ B {\bf 602}, 52 (2004).


\bibitem{Pal:2003zf}
S.~Pal and S.~Pratt,
Phys.\ Lett.\ B {\bf 574}, 21 (2003).


\bibitem{Djordjevic:2003zk}
M.~Djordjevic and M.~Gyulassy,
Nucl.\ Phys.\ A {\bf 733}, 265 (2004).


\bibitem{alphas}
O.~Kaczmarek {\it et al.},  
hep-lat/0406036;
F. Znatow, these proceedings. 


\bibitem{Collins:1981uk}
J.~C.~Collins and D.~E.~Soper,
Nucl.\ Phys.\ B {\bf 193}, 381 (1981)
[Erratum-ibid.\ B {\bf 213}, 545 (1983)].



\bibitem{jets}
H.~Takai, 
J.\ Phys.\ G {\bf 30}, S1105 (2004); these proceedings; 
J. Liu, these proceedings. 



\end{thebibliography}
\end{document}